\documentclass[12pt,a4paper]{article}
\usepackage{bbm}
\usepackage{amssymb,amsmath}
\usepackage{rotating}
\usepackage[centerlast]{caption}
\usepackage{graphicx}
\usepackage{verbatim}
\usepackage{enumerate}
\usepackage{mathrsfs}
\usepackage{amsfonts}
\usepackage{multirow}
\usepackage{amsthm,amscd,graphics,latexsym,color}
\usepackage{booktabs,tabularx}
\usepackage{caption}
\usepackage{subcaption}
\usepackage[title,titletoc]{appendix}
\usepackage[authoryear]{natbib}
\usepackage{amsmath}
\usepackage{amssymb}

\usepackage[a4paper,text={16.0cm, 25.0cm},centering]{geometry}

\def\be{\begin{equation}}
\def\ee{\end{equation}}
\def\bestar{\begin{eqnarray*}}
\def\eestar{\end{eqnarray*}}
\def\bea{\begin{eqnarray}}
\def\eea{\end{eqnarray}}
\theoremstyle{remark}

\theoremstyle{definition}

\theoremstyle{plain}
\newtheorem{theo}{Theorem}
\theoremstyle{plain}

\theoremstyle{plain}
\newtheorem{cor}{Corollary}
\theoremstyle{plain}

\theoremstyle{plain}
\newtheorem{lemma}{Lemma}
\newtheorem{example}{Example}

\numberwithin{equation}{section} \numberwithin{theo}{section}
\numberwithin{defn}{section} \numberwithin{rem}{section}
\numberwithin{cor}{section} \numberwithin{lemma}{section}
\numberwithin{prop}{section} \numberwithin{assumption}{section}
\allowdisplaybreaks[4]

\newcommand{\D}{\displaystyle}
\newcommand{\DF}[2]{\frac{\D#1}{\D#2}}
\newcommand{\e}{{\mathbb E}}
\newcommand{\p}{{\mathbb P}}
\newcommand{\ra}{{\cal F}}

\begin{document}

\begin{titlepage}

\begin{center}
{\bf \Large Smoothing the Nonsmoothness}

\bigskip

 {\sc Chaohua Dong$^{a}$, Jiti Gao$^{b}$, Bin Peng$^{b}$ and Yundong Tu$^{c}$}
 
 \medskip
 
 $^{a}$Zhongnan University of Economics and Law, China\\ $^{b}$Monash University, Australia\\$^{c}$Peking University, China
\end{center}

\date{\today}
 
\begin{abstract}
To tackle difficulties for theoretical studies in situations involving nonsmooth functions, we propose a sequence of infinitely differentiable functions to approximate the nonsmooth function under consideration. A rate of approximation is established and an illustration of its application is then provided.

\bigskip

\end{abstract}

\end{titlepage}

\section{Introduction}\label{Sec1}

This paper proposes a unified smoothing technique that aims at providing a generic framework to tackle difficulties in theoretical analysis in the situations that involve nonsmooth functions. These situations mostly encountered are objective functions constructed by nonsmooth loss functions (e.g. least absolute deviation (LAD), quantile loss and Huber's loss) or approximations to some functions of interest using nonsmooth activation functions (e.g. ReLU) in machine learning. The relevant literature includes \citet{huber1964}, \cite{koenker1978}, \citet{bai1992}, \citet{fanjq1994}, \cite{Tianetal}, \citet{xiao2009a}, \citet{fan2018}, \citet{degui2021}, \citet{fernandes2021}, \citet{zhouwx2022}, \citet{hexm2023},  just to name a few. In what follows, we give two well known examples to illustrate our concerns.
\smallskip

\begin{example} \label{EX1}
Consider the rectified linear unit (ReLU)
\begin{eqnarray}
\rho(x) = x\vee 0,
\end{eqnarray}
which is widely adopted for different types of deep neural network studies. However, ReLU is piecewise linear, and it is not yet clear how to handle the accumulated non-differentiability through layers in both theory and practice. The concern raised here actually exists in some well known software packages. To the best of our knowledge, however, no satisfactory treatment has been offered. For example, the well known \texttt{neuralnet} in R does not even support the use of ReLU (\citealp{FG2010}). \texttt{PyTorch} does have ReLU and some of its variations included as the activation functions, but the explanation about the optimization process is very vague (https://pytorch.org/docs/stable/optim.html). \texttt{Keras} includes \texttt{Adam} algorithm and its variations (https://keras.io/api/optimizers/adam/), but \texttt{Adam} requires ``\textit{a stochastic scalar function that is differentiable w.r.t. parameters...}'' (\citealp{KingBa15}), which clearly does not apply to ReLU.
\end{example}
\smallskip

\begin{example}\label{EX2}
Consider a regression where an objective function is constructed using a nonsmooth function loss $\rho(\cdot)$ say to estimate unknown parameter. The most popular loss is the check function. Nonsmoothness of $\rho(\cdot)$ gives rise to difficulties in both theoretical analysis and computation especially  when handling large datasets (\citealp{hexm2023}). As can be seen, \cite{hexm2023} propose a convolution-type smoothed quantile loss to replace the usual check function.
\end{example}
\smallskip

In this paper we are about to provide a unified approach through generalized functions to tackle a class of nonsmooth functions in statistics and machine learning. The class includes several commonly encountered losses such as quantile loss, LAD, Huber's loss and ReLU. Also, our treatment is independent of data that is advantageous over existing smoothing methods. Last but not least, we are able to establish a rate of convergence for a sequence of infinitely differentiable functions that can converge to the nonsmooth function. It is our knowledge that the relevant literature (see, for example, \citealt{phillips1991, phillips1995}) has not established such a rate of convergence. Because of this approximation rate the theoretical analysis that we shall show in Section \ref{Sec3} below is in a solid ground.

Often, nonsmooth losses have second derivatives in generalized function sense (e.g. Dirac delta function) that however cannot be used directly in theoretical analysis. Nonetheless, their operation is valid under integration with certain conditions. For example, for any random variable $e$ with density function $f_e(x)$, $\e[\delta(e)]=\int \delta(x)f_e(x)dx=f_e(0)$, provided $f_e(x)$ is continuous at $x=0$, where $\delta(\cdot)$ is Dirac delta function. This motivates us to use a sequence of infinitely differentiable functions (regular sequence) to approximate any given nonsmooth function, and the derivatives of the regular sequence converge to the derivatives of the nonsmooth function in generalized function sense, so that in some suitable circumstance the nonsmooth function under consideration can be replaced by the regular sequence. This is the rationale of our approach.

In the sequel, Section 2 shows the smoothing method we propose, Section 3 gives one of its applications, and Monte Carlo simulations and all proofs of our theoretical results are given in an online material.

\section{Generalized functions and smoothing technique}\label{Sec2}

This section will show the rationale of the proposed smoothing technique based on a generalized function approach; we shall define a regular sequence for a given nonsmooth function, and  establish its rate of convergence.

The following assumption describes a class of nonsmooth functions $\rho(\cdot)$ that we work with. As $\rho(\cdot)$ is convex, its subgradient $\psi(\cdot)$ always exists even though the value of $\psi(\cdot)$ at some points may not be unique; we allow $\psi(\cdot)$ to take any of them while the analysis in the sequel remains unchanged. The focus then rests on the second derivative of $\rho(\cdot)$ that may not exist in ordinary sense.

\smallskip

\noindent \textbf{Assumption 2.1}.
\begin{enumerate}[(a)]
\item \emph{Suppose that $\rho(\cdot)$ is positive and convex, and has its only minimum at zero; moreover, it is a locally integrable function on the real line with the rate of diverging at infinity no faster than that of a polynomial.}
\item \emph{$\rho(\cdot)$ satisfies Lipschitz condition $|\rho(x+u)-\rho(u)|\le C |x|$ for any $x, u\in \mathbb{R}$, where $C$ is an absolute constant}.
\end{enumerate}

By convexity, $\rho(\cdot)$ is a continuous function (see, for example, Corollary 10.1.1 of \citet[p 83]{rockafellar1970}). The other conditions in (a) ensure that the proposed generalized function approach is valid (see the definition of space $S$ in supplementary material), and these are certainly fulfilled by all loss functions encountered in the literature and ReLU. The Lipschitz condition in (b) plays a central role that depicts a technical requirement in the following analysis, and it is guaranteed by the boundedness of subgradient. Note also that it is satisfied by several most popular loss functions and ReLU. For example, when $\rho(u)=|u|$, we have $C=1$; when $\rho(u)$ is the check function with parameter $\tau\in (0,1)$, we have $C=\max(\tau, 1-\tau)$; when $\rho(u)$ is Huber's loss with parameter $c$, we have $C=c$; for ReLU $\rho(u)=\max(0, u)$, $C=1$.

To define the regular sequence of loss functions we need the following assumption.

\smallskip

\noindent \textbf{Assumption 2.2}.
\begin{enumerate}[(a)]
\item \emph{Suppose that $\phi(\cdot)$ is nonnegative, symmetric and $\int \phi(u)dv=1$.}
\item \emph{Suppose that} $\phi(u)$ \emph{is infinitely differentiable, and $|u|^k\phi^{(\ell)}(u)\to 0$ for any nonnegative integers $k, \ell$ as $|u|$ approaches the boundary of the domain of $\phi(\cdot)$.}
\end{enumerate}
It is readily seen that any smooth and thin-tailed density satisfies this assumption, while the most popular one is the normal density. In addition, a bounded supported density function satisfies this assumption easily as long as it is infinitely differentiable. For example,
\begin{equation}\label{b1}
\phi(u)=\begin{cases}C\exp\left(-\DF{1}{1-u^2}\right), & |u|<1\\
0, & |u|\ge 1.
\end{cases}
\end{equation}
where $C$ is chosen to make the integration of $\phi(u)$ to be one. Note that the support of $\phi(u)$ can be changed to be any symmetric interval that however is not essential, and this assumption excludes densities such as uniform densities since they are not smooth on the real line. In addition, Assumption 2.2(b) ensures existence of all moments of $\phi(u)$, and coupled with Assumption 2.1, Assumption 2.2 (b) implies $\rho(u)\phi(u)\to 0$ as $|u|$ tends to the boundary of the domain of $\phi(\cdot)$.

\smallskip

\noindent \textbf{Definition 2.1}
\smallskip

\emph{Let $\rho(\cdot)$ and $\phi(\cdot)$ satisfy Assumptions 2.1 and 2.2, respectively. For a given positive integer $m$, define $\phi_m(u)=m\phi(mu)$, and}
\begin{equation}\label{b2}
\rho_m(u)=\int \rho(x)\phi_m(x-u)dx.
\end{equation}
\emph{The function sequence $\{\rho_m(u)\}$ is called a regular sequence of $\rho(u)$.}

\smallskip

Next, we establish some crucial results about a rate of convergence for a regular sequence of nonsmooth functions converging to the original nonsmooth function under consideration, which is used as a bridge between nonsmooth loss and its infinitely smooth approximation counterpart in the generalized function context.

\medskip

\begin{theo}\label{th21}
Let $\rho(u)$ and $\phi(u)$ satisfy Assumptions 2.1 and 2.2, respectively, and  $\rho_m(u)$ be the regular sequence of $\rho(u)$ given by \eqref{b2}. Then, (1) $\rho_m(u), m=1,2,\cdots,$ are convex;

(2) $\rho_m(u), m=1,2,\cdots,$ are differentiable with any order, and in particular,
\begin{align}\label{derivatives2a}
  \rho_m'(u)=&-\int \rho(x)\phi_m'(x-u)dx, & \rho_m''(u)=&\int \rho(x)\phi_m''(x-u)dx,
\end{align}
and $\rho_m'(u)\to \psi(u)$ and $\rho_m''(u)\to \rho''(u)$ as $m\to\infty$ except on a set of measure zero;

(3) $\sup_{u\in \mathbb{R}}|\rho_m(u)-\rho(u)|\le Cm^{-1}$ for some absolute constant $C$.
\end{theo}

\medskip

The proof is given in Appendix.
\smallskip

\noindent{\bf Remark 2.1}. \ Note that the assertion (1) confirms that all functions in the regular sequence of $\rho(u)$ remain convexity as $\rho(u)$ does. The assertion (2) is understandable in generalized function sense because $\phi_m(u)$ plays a role as a delta-convergent sequence, i.e. $\phi_m(x)\to \delta(x)$ as $m\to \infty$; see \citet{kanwal1983} and \citet{stein2003}.
\smallskip

\begin{figure}[h]
  \centering
  \hspace*{-3cm}  \includegraphics[scale=0.2]{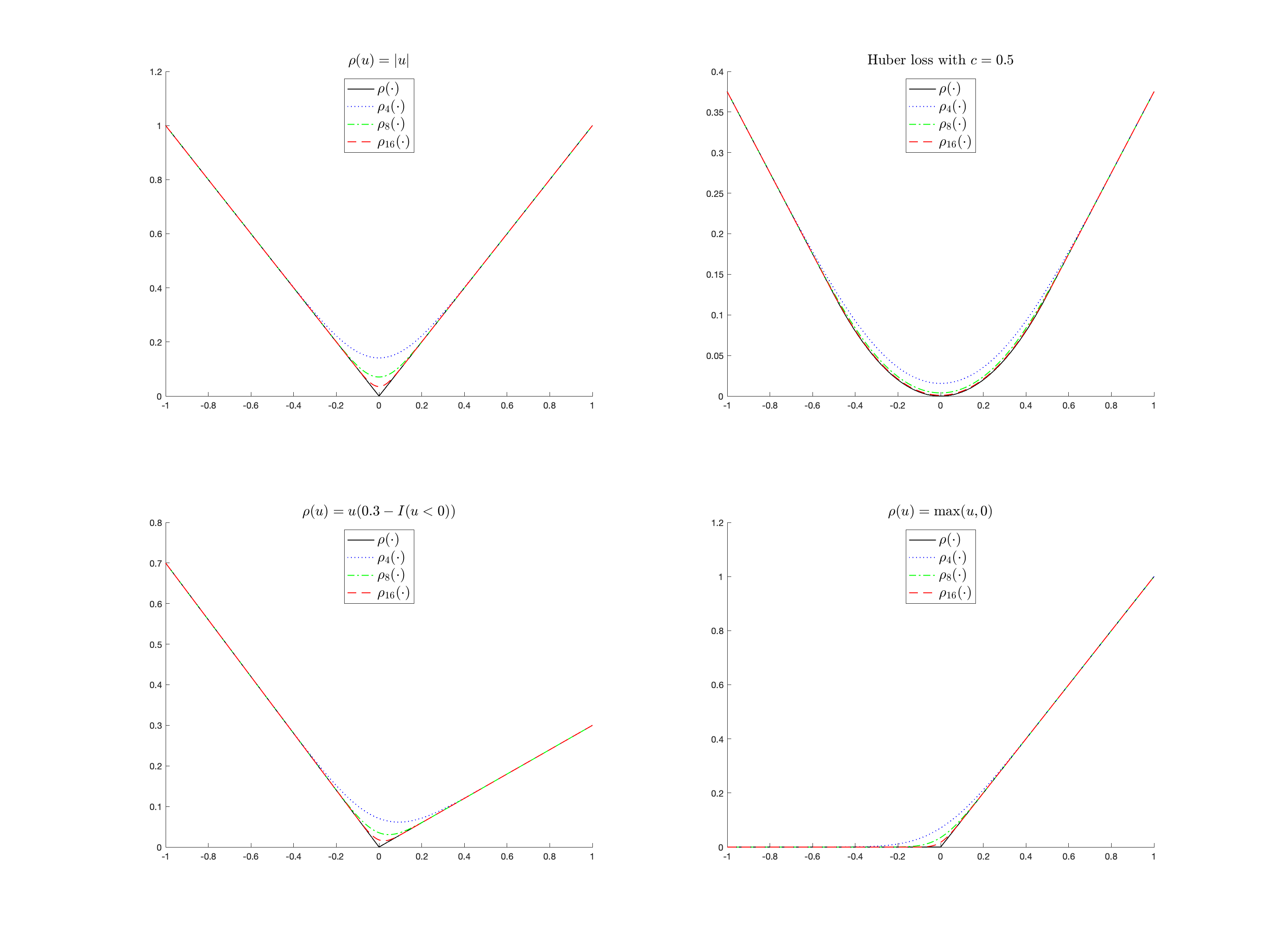}\\
  \caption{Regular sequences generated by normal density}\label{oldphi}
  \end{figure}

\noindent{\bf Remark 2.2}. \ The importance of Theorem \ref{th21} is the rate of convergence: $\sup_{u\in \mathbb{R}}|\rho_m(u)-\rho(u)|\le Cm^{-1}$. To the best of our knowledge, the rate is the first one being established in the relevant literature. It is due to this rate that our analysis below is on a solid and rigorous ground. In Figure \ref{oldphi}, three losses and ReLU along with their regular sequences generated by standard normal density are plotted to visualize the approximations established in Theorem \ref{th21} where, as can be seen, all kinks are smoothed and the regular sequences approach the respective functions well. Note also that the Lipschitz condition in Assumption 2.1 can be relaxed as $|\rho(x)-\rho(y)|\le C|x-y|^\alpha$ for $\alpha\in (0,1]$. All the results below under the relaxation hold with a change on the choice of $m$. \qed

For these particular nonsmooth functions and $\phi(u)$ given in \eqref{b1}, we have the following results that improve Theorem \ref{th21}.

\medskip

\begin{cor}\label{cor21}\it
Suppose that $\rho_m(u)$ below is generated by $\phi(u)$ given in \eqref{b1}. Let $\epsilon>0$ be given arbitrarily.
\begin{enumerate}[(a)]
  \item Let $\rho(u)=|u|$, $\rho(u)=u(\tau-I(u<0))$, or $\rho(u)=0\vee u$. There exists an integer $m_0=m_0(\epsilon)$ such that $\rho_m(u)=\rho(u)$ for all $m\ge m_0$ and all $|u|>\epsilon$, whereas (2) of Theorem \ref{th21} remains true for $|u|\le \epsilon$, viz. $\sup_{|u|\le \epsilon}|\rho_m(u)-\rho(u)|=O(m^{-1})$ as $m\to \infty$.
  \item For Huber's loss, $\rho_c(u)$, there exists an integer $m_0=m_0(\epsilon)$ such that $\rho_m(u)=\rho(u)$ for all $m\ge m_0$ and $|u|>c+\epsilon$, whereas (2) of Theorem \ref{th21} is improved, viz. $\sup_{|u|\le c}|\rho_m(u)-\rho(u)|=O(m^{-2})$ as $m\to \infty$.
\end{enumerate}
\end{cor}

\medskip

\begin{figure}[h]
\centering
 \hspace*{-3cm}  \includegraphics[scale=0.2]{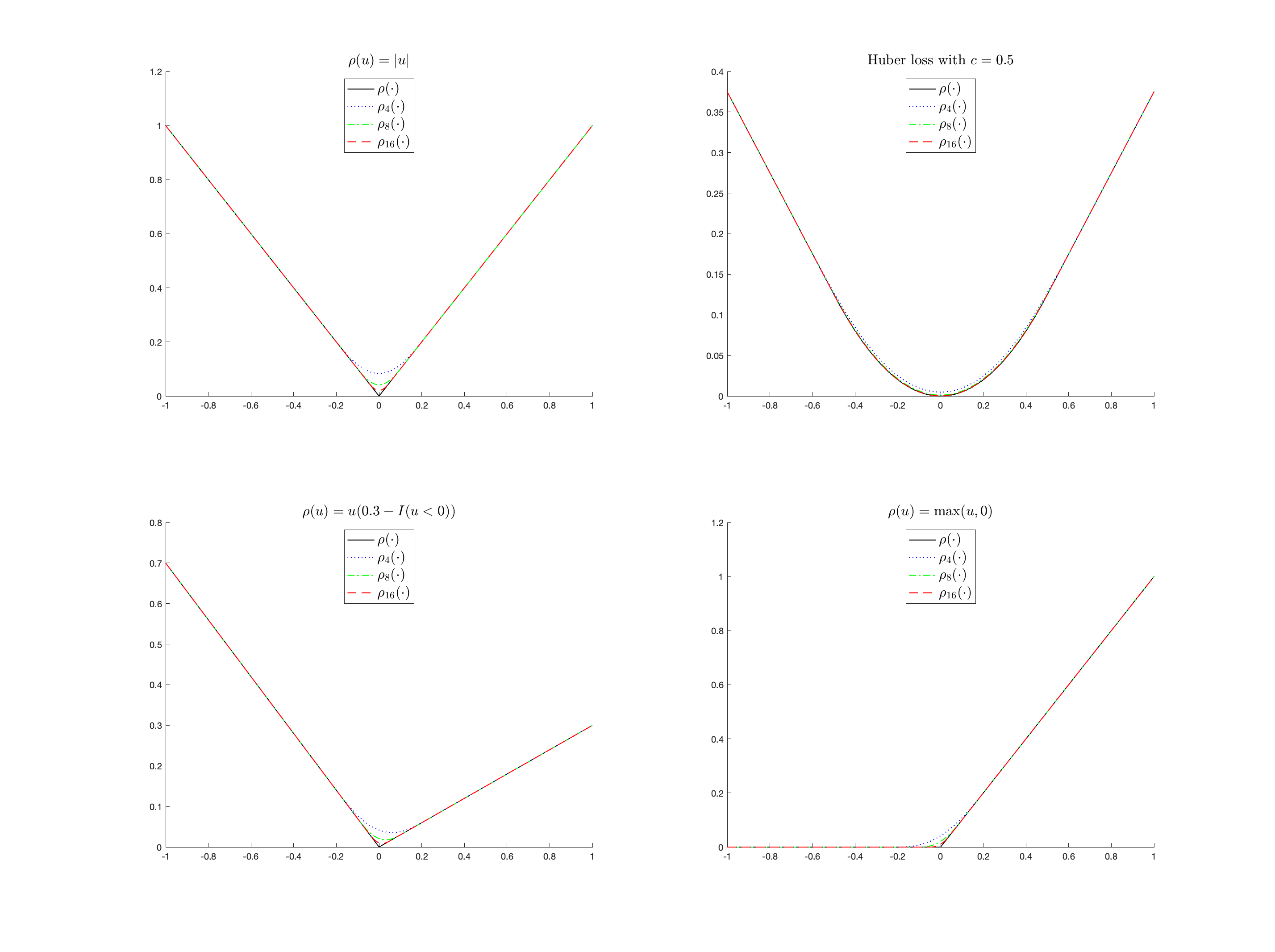}\\
  \caption{Regular sequences using $\phi(u)$ given by \eqref{b1}}\label{newphi}
  \end{figure}

The proof of this corollary is given in Appendix. Because of compactness of the domain of $\phi(\cdot)$, for these nonsmooth functions the corollary claims $\rho_m(u)=\rho(u)$ for $|u|>\epsilon$ (or $|u|>c+\epsilon$) uniformly for all $m\ge m_0(\epsilon)$ for given $\epsilon>0$; also for Huber's loss the convergence rate has been improved to be $m^{-2}$ in $|u|<c$. Figure \ref{newphi} shows the plots of the three loss functions and ReLU with their regular sequences constructed from $\phi(u)$ in \eqref{b1}. Apparently, Figure \ref{newphi} shows better approximation.

\section{Application of the smoothing technique}\label{Sec3}

To show the essence of the proposed smoothing technique, in this section we shall illustrate through a linear parametric regression. To do so, let us consider a linear regression of the form:
\begin{align}
y_i=&x_i^{\prime} \theta_0+e_i, \ \ \ i=1, 2, \cdots, n, \label{2a}
\end{align}
where the unknown parameter $\theta_0\in \Theta$, a compact subset of $\mathbb{R}^d$. The estimator of $\theta_0$ in the regression \eqref{2a} is defined by
\begin{equation}\label{2b}
\widehat{\theta}=\underset{\theta\in \Theta}{\arg\min}\ L_n(\theta)=\sum_{t=1}^n\rho(y_t- x_{t}'\theta),
\end{equation}
where $\rho(\cdot)$ satisfies Assumption 2.1.

To analyse $\widehat{\theta}$, instead of considering minimization over $\theta\in\Theta$, we shall follow the literature to consider vectors $\beta$ in the tangent cone $T_{\Theta}(\theta_0)$ of $\Theta$ at $\theta_0$, that is, $\beta=\sqrt{n}(\theta-\theta_0)$ where $\theta\in\Theta$. The relevant literature includes \citet{bickel1974}, \citet{badu1989}, \citet{davis1992}, \citet{charles1994}, \citet{phillips1995} and \cite{gao2009}.

We then focus on the minimization of $L_n(\theta)-L_n(\theta_0)$ instead of $L_n(\theta)$ in \eqref{2b}, and write
\begin{align*}
L_n(\theta)-L_n(\theta_0)
=&\sum_{t=1}^n[\rho(e_t-n^{-1/2}x_t'\beta)-\rho(e_t)]
\equiv\tilde{L}_n(\beta).
\end{align*}
Hereby, the objective function is reparametrized in $\beta\in \mathbb{R}^d$, and when $\widehat\beta$ minimizes $\tilde{L}_n(\beta)$, $\widehat\theta$ minimizes $L_n(\theta)$ where $\widehat\beta=\sqrt{n}(\widehat\theta -\theta_0)$.
\medskip

\begin{theo}\label{th22} \it
Suppose that $\{e_i, x_i\}_{i=1}^n$ is an independent and identically distributed (i.i.d.) sequence, $e_i$ and $x_i$ are mutually independent, and the fourth moment of $x_i$ exists. Define
\begin{equation}\label{th22a}
Q_n(\beta)=\left(-\DF{1}{\sqrt{n}}\sum_{i=1}^n\psi(e_i)x_i'\right)\beta
+\DF{a}{2}\beta' \left(\DF{1}{n} \sum_{t=1}^nx_ix_i'\right)\beta,
\end{equation}
where $a=\e[\rho''(e_i)]>0$. Then,
\begin{equation}\label{th22b}
  |\tilde{L}_{n}(\beta)-Q_n(\beta)|=O_P(n^{-1/2}),
\end{equation}
for each $\beta$; and
\begin{equation}\label{th22c}
  \sup_{\|\beta\|\le c}|\tilde{L}_{n}(\beta)-Q_n(\beta)|=O_P(n^{-1/2}),
\end{equation}
for any $c>0$; moreover,
\begin{equation}\label{th22d}
  \sup_{\|\beta\|\le c_n}|\tilde{L}_{n}(\beta)-Q_n(\beta)|=o_P(1),
\end{equation}
for any $c_n=o(n^{1/4})$.

Furthermore, define $\widehat\beta_Q$ as the minimizer of $Q_n(\beta)$, and suppose that $\e[\psi(e_i)]=0$ and the minimum eigenvalue of $\DF{1}{n} \sum_{t=1}^nx_ix_i'$, $\lambda_{\min}>0$, for all large $n$. Then as $n\to\infty$ we have
\begin{equation}\label{th22e}
\widehat{\beta}-\widehat\beta_Q=o_P(n^{-1/4}\log\log(n)).
\end{equation}
\end{theo}

The proof is regelated to Appendix. Theorem \ref{th22} proposes a quadratic form $Q_n(\beta)$ and gives several approximation errors of $\tilde{L}_{n}(\beta)-Q_n(\beta)$ as well as the error between their minimizers. It is clear that
\begin{align*}
\widehat{\beta}_Q=\left(\DF{a}{n} \sum_{i=1}^nx_ix_i'\right)^{-1} \left(\DF{1}{\sqrt{n}}\sum_{i=1}^n\psi(e_i)x_i\right),
\end{align*}
and it is the unique minimizer of $Q_n(\beta)$.
\smallskip

\noindent{\bf Remark 3.1}. \ As argued in \citet[p187]{pollard1991}, quadratic approximation to objective function of M estimation avoids technical difficulty, such as stochastic equicontinuity in an asymptotic proof. Indeed, equation \eqref{th22e} gives $\widehat{\beta}=\widehat\beta_Q+o_P(n^{-1/4}\log\log(n))$. Once one has the limit of $\widehat\beta_Q$ after specifying $\rho(\cdot)$, the limit of $\widehat\beta=\sqrt{n}(\widehat\theta -\theta_0)$ is available immediately that is the main purpose of the analysis for the estimator derived from the loss $\rho(\cdot)$. For example, if $\rho(u)=u(\tau-I(u<0))$, then $a=f_e(0)$, $\e[\psi(e_i)]^2=\e[\tau-I(e_i<0)]^2=\tau(1-\tau)$; suppose moreover $\DF{1}{n} \sum_{i=1}^nx_ix_i'\to_P\Sigma$, then $\widehat\beta_Q\to_D[\sqrt{\tau(1-\tau)}/f_e(0)]N(0, \Sigma^{-1})$ as $n\to\infty$ that is also the limit of $\sqrt{n}(\widehat\theta -\theta_0)$.
\smallskip

\noindent{\bf Remark 3.2}. \ \citet[p308]{fan2003} use quadratic approximation for smooth objective function and derive limit theory from this quadratic function. By contrast, the generalized function approach we propose mainly deals with nonsmooth loss functions.

\section{Conclusion}

In this paper, we propose a smoothing technique providing convenient and efficient tool to tackle difficulties in theoretical analysis in the situations that involve nonsmooth functions. Precisely, an infinitely differentiable sequence that converges to the nonsmooth function is proposed; convergence rate is established that facilitates theoretical analysis. An application is provided to illustrate the usefulness of the smoothing technique.

%\section*{Acknowledgements}
%
%We would like to thank all participants of the 21st Symposium of Statistic Science of China, Peking University, and 2022 Modern Statistics Symposium, Xiamen University. Also, we thank Mervyn Silvapulle and Roufan Xu for their constructive comments and suggestions.
%
%Dong would like to thank the financial support from National Natural Science Foundation of China (Grant 72073143) and the Fundamental Research Funds for the Central Universities, Zhongnan University of Economics and Law (2722022EG001); Gao and Peng acknowledge financial support from the Australian Research Council Discovery Grants Program under Grant Numbers: DP200102769 \& DP210100476, respectively; and Tu (Corresponding Author) would like to thank support from National Natural Science Foundation of China (Grant 72073002, 12026607, 92046021), the Center for Statistical Science at Peking University, and Key Laboratory of Mathematical Economics and Quantitative Finance (Peking University), Ministry of Education.

\bibliographystyle{biometrika}
\bibliography{paper-ref}

\begin{thebibliography}{}

\bibitem[Badu, 1989]{badu1989}
Badu, G.~J. (1989).
\newblock {Strong representations for LAD estimators in linear models}.
\newblock {\em Probability theory and related fields}, 83:547--558.

\bibitem[Bai et~al., 1992]{bai1992}
Bai, Z., Rao, C.~R., and Wu, Y. (1992).
\newblock {M estimation of multivariate linear regression parameters under a
  convex discrepancy function}.
\newblock {\em Statistica Sinica}, 2:237--254.

\bibitem[Bickel, 1974]{bickel1974}
Bickel, P.~J. (1974).
\newblock {Edgeworth expansions in nonparametric statistics}.
\newblock {\em Annals of Statistics}, 2:1--20.

\bibitem[Bravo et~al., 2021]{degui2021}
Bravo, F., Li, D., and Tj{\o}stheim, D. (2021).
\newblock {Robust nonlinear regression estimation in null recurrent time
  series}.
\newblock {\em Journal of Econometrics}, 224:416--438.

\bibitem[Davies et~al., 1992]{davis1992}
Davies, R.~A., Knight, K., and Liu, J. (1992).
\newblock {M-estimation for autoregressions with infinite variance}.
\newblock {\em Stochastic Processes and Their Application}, 40:145--180.

\bibitem[Fan et~al., 1994]{fanjq1994}
Fan, J., Hu, T.-C., and Truong, Y.~K. (1994).
\newblock {Robust nonparametric function estimation}.
\newblock {\em Scandinavian Journal of Statistics}, 21:433--446.

\bibitem[Fan et~al., 2003]{fan2003}
Fan, J., Jiang, J., Zhang, C., and Zhou, Z. (2003).
\newblock {Time-dependent diffusion models for term structure dynamics}.
\newblock {\em Statistica Sinica}, 13:965--992.

\bibitem[Fernandes et~al., 2021]{fernandes2021}
Fernandes, M., Guerre, E., and Horta, E. (2021).
\newblock Smoothing quantile regressions.
\newblock {\em Journal of Business and Economic Statistics}, 39(1):338--357.

\bibitem[Gao et~al., 2009]{gao2009}
Gao, J., Li, D., and Lin, Z. (2009).
\newblock Robust estimation in parametric time series models under long-- and
  short--range dependent structures.
\newblock {\em Australian and New Zealand Journal of Statistics},
  51(2):161--181.

\bibitem[Geyer, 1994]{charles1994}
Geyer, C.~J. (1994).
\newblock {On the asymptotics of constrained $M$-estimation}.
\newblock {\em The Annals of Statistics}, 22:1993--2010.

\bibitem[G{\"u}nther and Fritsch, 2010]{FG2010}
G{\"u}nther, F. and Fritsch, S. (2010).
\newblock Neuralnet: {T}raining of neural networks.
\newblock {\em R Journal}, 2:30--38.

\bibitem[He et~al., 2023]{hexm2023}
He, X., Pan, X., Tan, K.~M., and Zhou, W.-X. (2023).
\newblock Smoothed quantile regression with large-scale inference.
\newblock {\em Journal of Econometrics}, 232:367--388.

\bibitem[Huber, 1964]{huber1964}
Huber, P.~J. (1964).
\newblock {Robust estimation of a location parameter}.
\newblock {\em Annals of Mathematical Statistics}, 35:73--101.

\bibitem[Kanwal, 1983]{kanwal1983}
Kanwal, R.~P. (1983).
\newblock {\em {Generalized Fuctions: Theory and Technique}}.
\newblock Academic Press, New York.

\bibitem[Kingma and Ba, 2015]{KingBa15}
Kingma, D. and Ba, J. (2015).
\newblock Adam: A method for stochastic optimization.
\newblock In {\em International Conference on Learning Representations (ICLR)},
  San Diega, CA, USA.

\bibitem[Koenker and Bassett, 1978]{koenker1978}
Koenker, R. and Bassett, G. (1978).
\newblock {Regression quantiles}.
\newblock {\em Econometrica}, 46:33--50.

\bibitem[Phillips, 1991]{phillips1991}
Phillips, P. C.~B. (1991).
\newblock {A shortcut to LAD estimator asymptotics}.
\newblock {\em Econometric Theory}, 7:450--463.

\bibitem[Phillips, 1995]{phillips1995}
Phillips, P. C.~B. (1995).
\newblock {Robust nonstationary regression}.
\newblock {\em Econometric Theory}, 11:912--951.

\bibitem[Pollard, 1991]{pollard1991}
Pollard, D. (1991).
\newblock {Asymptotics for the least absolute deviation regression estimators}.
\newblock {\em Econometric Theory}, 7:186--199.

\bibitem[Rockafellar, 1970]{rockafellar1970}
Rockafellar, R.~T. (1970).
\newblock {\em {Convex Analysis}}.
\newblock Princeton University Press, Princeton.

\bibitem[Rudin, 2004]{rudin2004}
Rudin, W. (2004).
\newblock {\em {Principles of Mathematical Analysis}}.
\newblock McGraw-Hill Companies, Inc., New York.

\bibitem[Stein and Shakarchi, 2003]{stein2003}
Stein, E.~M. and Shakarchi, R. (2003).
\newblock {\em {Fourier Analysis An Introduction}}.
\newblock Princeton University Press, Princeton.

\bibitem[Tan et~al., 2022]{zhouwx2022}
Tan, K., Wang, L., and Zhou, W.-X. (2022).
\newblock High-dimensional quantile regression: Convolution smoothing and
  concave regularization.
\newblock {\em Journal of the Royal Statistical Society: Series B (Statistical
  Methodology)}, 84:205--233.

\bibitem[Tian et~al., 2004]{Tianetal}
Tian, L., Liu, J., Zhao, Y., and Wei, L.~J. (2004).
\newblock Statistical inference based on non-smooth estimating functions.
\newblock {\em Biometrika}, 91(4):943--954.

\bibitem[Xiao, 2009]{xiao2009a}
Xiao, Z. (2009).
\newblock Quantile cointegrating regression.
\newblock {\em Journal of Econometrics}, 150:248--260.

\bibitem[Zhou et~al., 2018]{fan2018}
Zhou, W., Bose, K., Fan, J., and Liu, H. (2018).
\newblock {A new perspective on robust M-estimation: finite sample theory and
  applications to dependence-adjusted multiple testing}.
\newblock {\em Annals of Statistics}, 46:1904--1931.

\end{thebibliography}

{\small

\newpage
\begin{center}
\large \bf Supplementary material to \\``Smoothing the Nonsmoothness''
\end{center}

\begin{center}
Chaohua Dong$^{a}$, Jiti Gao$^{b}$, Bin Peng$^{b}$ and Yundong Tu$^{c}$\\
$^{a}$Zhongnan University of Economics and Law, China\\ $^{b}$Monash University, Australia\\$^{c}$Peking University, China
\end{center}

The supplementary material provides two sections, Monte Carlo simulations and the proofs of the theorems and corollary.

\appendix

\section{Simulation experiments}

\renewcommand{\theequation}{A.\arabic{equation}}
\renewcommand{\thefigure}{A.\arabic{figure}}
\renewcommand{\thetable}{A.\arabic{table}}
\renewcommand{\thelemma}{A.\arabic{lemma}}

\setcounter{equation}{0}
\setcounter{lemma}{0}
\setcounter{table}{0}
\setcounter{figure}{0}

In what follows, we use some simulated data to examine the finite sample performance of the newly proposed method through a quantile regression,
\begin{eqnarray}\label{simulation1}
y_i = x_i\theta_0+e_i, \ \ i=1,\ldots, n,
\end{eqnarray}
where $x_i$ is scalar and $x_i\sim N(1,1)$, $\theta_0=1$, $e_i =\varepsilon_i - F_\varepsilon^{-1}(\tau)$. We consider two cases for $\varepsilon_i$: \textbf{Thin tail}: $\varepsilon_i\sim N(0,1)$; and \textbf{Heavy tail}: $\varepsilon_i\sim t_4$, where $t_4$ stands for $t$-distribution with degree of freedom 4.

For each generated dataset, we carry on a quantile regression\footnote{The Matlab program is obtained from  http://www.econ.uiuc.edu/~roger/research/rq/rq.html.} to obtain $\widehat{\theta}_{\tau}.$  Also, we obtain minimizers $\widehat{\theta}_m$ with $m=5,\ 10,\ 15$ by replacing $\rho(\cdot)$ with $\rho_m(\cdot)$ in the objective function (the same for $\widehat{\beta}_m$ below), and $\phi(\cdot)$ with compact support is used in the construction of $\rho_m(\cdot)$. For the purpose of comparison, we consider the convolution  smoothed approach of \cite{hexm2023} to obtain $\widehat{\theta}_h$ with $ h=0.1,\ 0.5, \ 0.9.$ In their work, $h\in(0,1)$, so the above choices do not loss any generality. After $M$ replications, we calculate
\begin{eqnarray*}
\text{RMSE}_{\tau} &=&\left\{ \frac{1}{M}\sum_{j=1}^M (\widehat{\theta}_{\tau,j} -\theta_0)^2 \right\}^{1/2},\quad
\text{RMSE}_m =\left\{ \frac{1}{M}\sum_{j=1}^M (\widehat{\theta}_{m,j} -\theta_0)^2 \right\}^{1/2},\nonumber \\
\text{RMSE}_h &=&\left\{ \frac{1}{M}\sum_{j=1}^M (\widehat{\theta}_{h,j} -\theta_0)^2 \right\}^{1/2},
\end{eqnarray*}
where $\widehat{\theta}_{\tau,j}$, $\widehat{\theta}_{m,j}$ and $\widehat{\theta}_{h,j}$ are the values of $\widehat{\theta}_{\tau}$, $\widehat{\theta}_{m}$ and $\widehat{\theta}_{h}$ at the $j^{th}$ replication. Without loss of generality, we let $M=1000$, $\tau =0.3, 0.7$, and $n=100, 200$.

\begin{table}[h]
\centering
\caption{Root Mean Squared Errors (RMSEs)}\label{T1}
\begin{tabular}{llrrrrr}
\hline\hline
 &  & \multicolumn{2}{c}{$\tau =0.3$} & \multicolumn{1}{l}{} & \multicolumn{2}{c}{$\tau =0.7$} \\
 &  & $n=100$ & $n=200$ &  & $n=100$ & $n=200$ \\ \hline
 &  & \multicolumn{5}{c}{Heavy Tail: $t_4$} \\
\multicolumn{2}{l}{$\text{RMSE}_{\tau}$} & 0.106 & 0.298 &  & 0.106 & 0.303 \\
\multicolumn{1}{c}{\multirow{3}{*}{$\text{RMSE}_m$}} & $m=5$ & 0.103 & 0.298 &  & 0.103 & 0.302 \\
 & $m=10$ & 0.110 & 0.298 &  & 0.105 & 0.303 \\
 & $m=15$ & 0.111 & 0.298 &  & 0.105 & 0.303 \\
\multicolumn{1}{c}{\multirow{3}{*}{$\text{RMSE}_h$}} & $h=0.1$ & 0.103 & 0.298 &  & 0.102 & 0.302 \\
 & $h=0.5$ & 0.105 & 0.295 &  & 0.102 & 0.299 \\
 & $h=0.9$ & 0.144 & 0.294 &  & 0.138 & 0.297 \\ \hline
  &  & \multicolumn{5}{c}{Thin Tail: $N(0,1)$} \\
\multicolumn{2}{l}{$\text{RMSE}_{\tau}$} & 0.086 & 0.341 &  & 0.085 & 0.342 \\
\multicolumn{1}{c}{\multirow{3}{*}{$\text{RMSE}_m$}} & $m=5$  & 0.083 & 0.341 &  & 0.081 & 0.341 \\
 & $m=10$  & 0.084 & 0.342 &  & 0.083 & 0.342 \\
 & $m=15$  & 0.085 & 0.342 &  & 0.083 & 0.342 \\
\multicolumn{1}{c}{\multirow{3}{*}{$\text{RMSE}_h$}} & $h=0.1$ & 0.082 & 0.341 &  & 0.081 & 0.341 \\
 & $h=0.5$ & 0.083 & 0.340 &  & 0.081 & 0.339 \\
 & $h=0.9$ & 0.136 & 0.340 &  & 0.134 & 0.339 \\
 \hline\hline
\end{tabular}
\end{table}

Note that the simulation experiment is not to show our smoothing method can outperform the existing ones in terms of RMSEs. Our goal is to show that the newly proposed methods is not sensitive to the choice of $m$, while it still offers the equivalent finite sample performance. This can be explained by Corollary 1 where we have shown $\rho(u)=\rho_m(u)$ for all $|u|>1/m$.  Based on Table \ref{T1}, the results support our goal in general, and RMSE$_m$ is roughly the same as RMSE$_\tau$ in almost every case. For the convolution smoothed approach, RMSE$_h$ seems to be more sensitive to the choice of $h$ when $n$ is small.

\medskip

Next, we examine our result on $\widehat{\beta}-\widehat{\beta}_Q$ in Theorem 2. The data generating process is the same as \eqref{simulation1} with $\tau=0.5$, the median regression. Thus, $a=f_e(0)$, where $f_e$ stands for the density function of $e_i$. For each generated dataset, we calculate $\text{MAD}_m = \frac{1}{M}\sum_{j=1}^M  |\widehat{\beta}_{m,j}-\widehat{\beta}_{Q,j} |,$ where $\widehat{\beta}_{m,j} $ and $\widehat{\beta}_{Q,j}$ stand for $\widehat{\beta}_m$ and $\widehat{\beta}_{Q}$ at the $j^{th}$ replication, and $\widehat{\beta}_m=\sqrt{n}(\widehat{\theta}_m-\theta_0)$.

\begin{table}[h]
\centering
\caption{Mean Absolute Distance (MAD)}\label{T2}
\begin{tabular}{llrr}
\hline\hline
& $m$  & $n=100$ & $n=200$ \\
Heavy Tail: $t_4$  & 5 & 0.838 & 0.787 \\
 & 10 & 0.831 & 0.775 \\
 & 15 & 0.831 & 0.772 \\ \hline
Thin Tail: $N(0,1)$ & 5 & 0.702 & 0.697 \\
 & 10 & 0.691 & 0.682 \\
 & 15 & 0.688 & 0.680 \\
 \hline\hline
\end{tabular}
\end{table}

By Table \ref{T2}, the results of MAD support our theoretical assertion. In Theorem 2 we shown $\widehat{\beta}-\widehat{\beta}_Q=O_P(n^{-1/4}\log \log (n))$, where $\widehat{\beta}=\sqrt{n}(\widehat{\theta}-\theta_0)$. Note that the quantity $O_P(n^{-1/4}\log \log (n))$ is not convergence rate, whereas it gives how far the minimizer of the quadratic form proposed in Theorem 2 is from the normalized estimator of $\widehat{\theta}$. Hence, it is not surprising these numbers converge to 0 slowly.

\renewcommand{\theequation}{B.\arabic{equation}}
\renewcommand{\thefigure}{B.\arabic{figure}}
\renewcommand{\thetable}{B.\arabic{table}}
\renewcommand{\thelemma}{B.\arabic{lemma}}

\setcounter{equation}{0}
\setcounter{lemma}{0}
\setcounter{table}{0}
\setcounter{figure}{0}

\section{Proofs of the main results}\label{appb}

{\bf Proof of Theorem \ref{th21}} ~~(1) By the convexity of $\rho(\cdot)$, for any $\lambda\in [0,1]$ and $u,v\in \mathbb{R}$,
\begin{align*}
\rho_m(\lambda u+(1-\lambda)v)=&\int \rho(\lambda u+(1-\lambda)v+\DF{1}{m}x)\phi(x)dx\\
\le&\lambda\int \rho( u+\DF{1}{m}x)\phi(x)dx+(1-\lambda)\int \rho( v+\DF{1}{m}x)\phi(x)dx\\
=&\lambda\rho_m(u)+(1-\lambda)\rho_m(v).
\end{align*}

(2) By Theorem 9.42 of \citet{rudin2004}, we may take derivative under the integral, and by integration by parts, we have
\begin{align*}
\rho_m'(u)=&-\int \rho(x)\phi_m'(x-u)dx=\int \psi(u+x)\phi_m(x)dx\to \psi(u)
\end{align*}
as $m\to\infty$ whenever $u$ is a continuity of $\psi(\cdot)$, because $\phi_m(x)\to \delta(x)$ in generalized function sense; see \citet[p.62]{kanwal1983}. Since the set of noncontinuous points has measure zero, the assertion follows. For the second derivative, the same argument applies.

(3) Note that $\phi_m(x)=m\phi(mx)$, and
\begin{align*}
\sup_u|\rho_m(u)-\rho(u)|=&\sup_u \left|\int \rho(x)\phi_m(x-u)dx-\rho(u)\right|\\
\le&\sup_u\int |\rho(x+u)-\rho(u)|\phi_m(x)dx\\
\le&C\int |x|\phi_m(x)dx =\frac{C}{m}\int |x|\phi(x)dx.
\end{align*}

\qed

\noindent{\bf Proof of Corollary 2.1}. Note that for any $\rho(\cdot)$, one always can write
\begin{align*}
\rho_m(u)-\rho(u)= & \int \rho(v)\phi_m(v-u)dv-\rho(u)
=\int_{-1}^1 \left[\rho\left(u+\DF{v}{m}\right)-\rho(u)\right]\phi(v)dv.
\end{align*}

(a) Since proofs for all $\rho(u)$ in the corollary are similar, we only show the assertion for $\rho(u)=|u|$. For $|u|>\epsilon$, choose an integer $m_0$ such that $m^{-1}_0<\epsilon$. Then, for all $m\ge m_0$ and any $v\in [-1,1]$, $u+v/m$ has the same sign as $u$; thus, $|u+v/m|-|u|=sgn(u)\; v/m$. Accordingly,
\begin{align*}
\rho_m(u)-\rho(u)=&\int_{-1}^1 \left[\left|u+\DF{v}{m}\right|-|u|\right]\phi(v)dv=sgn(u)\DF{1}{m}\int_{-1}^1 v\phi(v)dv=0,
\end{align*}
since $\phi(v)$ is symmetric. The assertion follows evidently.

(b) For Huber's loss, $\rho_c(u)=0.5 u^2I(|u|\le c)+(c|u|-0.5c^2)I(|u|>c)$ with $c>0$. If $|u|>c+\epsilon$, we choose an integer $m_0>\epsilon^{-1}$. Then, for all $m>m_0$ and any $v\in [-1,1]$, $|u+v/m|>c$ and $u+v/m$ has the same sign as $u$. Thus, for $m>m_0$,
\begin{align*}
\rho_m(u)-\rho(u)=&c\int_{-1}^1 \left[\left|u+\DF{v}{m}\right|-|u|\right]\phi(v)dv
=sgn(u)\DF{c}{m}\int_{-1}^1 v\phi(v)dv=0.
\end{align*}

If $|u|<c$, we choose an integer $m_0$ such that for all $m>m_0$ and any $v\in [-1,1]$, $|u+v/m|<c$. Thus, for $m>m_0$,
\begin{align*}
\rho_m(u)-\rho(u)=&\DF{1}{2}\int_{-1}^1 \left[\left|u+\DF{v}{m}\right|^2-|u|^2\right]\phi(v)dv
=\DF{1}{2}\int_{-1}^1 \left[2u\DF{v}{m}+\DF{v^2}{m^2}\right]\phi(v)dv\\
=&\DF{1}{2m^2}\int_{-1}^1v^2\phi(v)dv =C \DF{1}{m^2}.
\end{align*}

In addition, for $u=c$,
\begin{align*}
\rho_m(c)-\rho(c)
=&\int_{0}^1 \left[c\left(c+\DF{v}{m}\right)-\DF{1}{2}c^2-\DF{1}{2}c^2\right]\phi(v)dv
+\int_{-1}^0 \left[\DF{1}{2}\left(c+\DF{v}{m}\right)^2-\DF{1}{2}c^2\right]\phi(v)dv\\
=&\DF{c}{m}\int_{0}^1v\phi(v)dv+\DF{c}{m}\int_{-1}^0v\phi(v)dv+ \DF{1}{2m^2}\int_{-1}^0v^2\phi(v)dv
=C\DF{1}{m^2}.
\end{align*}

For $u=-c$,
\begin{align*}
\rho_m(-c)-\rho(-c)=&\int_{-1}^0 \left[c\left(c-\DF{v}{m}\right)-\DF{1}{2}c^2 -\DF{1}{2}c^2\right]\phi(v)dv
+\int_{0}^1 \left[\DF{1}{2}\left(-c+\DF{v}{m}\right)^2-\DF{1}{2}c^2\right]\phi(v)dv\\
=&-\DF{c}{m}\int_{-1}^0v\phi(v)dv-\DF{c}{m}\int_{0}^1v\phi(v)dv+ \DF{1}{2m^2}\int_{0}^1v^2\phi(v)dv
=C\DF{1}{m^2}.
\end{align*}
\qed

To prove Theorem \ref{th22}, we need the following lemma.
\smallskip

\begin{lemma}\label{lemma1}
Let $e$ be any random variable with density function $f(x)$. Under the same conditions as Theorem \ref{th21}, we have
(1) $\rho_m(e)-\rho(e)=O(m^{-1})$ with probability 1. (2) Suppose that $f(u)$ is differentiable, $\int |f'(u)|du <\infty$, and $\rho(u)f(u)\to 0$ as $|u|\to\infty$, then $\rho_m'(e)-\psi(e)=O_P(m^{-1})$. (3) Suppose that $\e[\rho''(e)]$ exists. When $\rho''(u)$ does not exist, the expectation is considered in generalized function sense; suppose also $\int |f''(u)|du <\infty$, $\rho(u)f(u)\to 0$ and $\psi(u)f'(u)\to 0$ when $|u|\to\infty$. Then $\e[\rho_m''(e)-\rho''(e)]=O(m^{-1})$. (4)  $\e[\rho_m''(e+\epsilon) -\rho''_m(e)]=O(m^{-1}+\epsilon)$ for any given small $\epsilon$.
\end{lemma}
\smallskip

\noindent{\bf Proof of Lemma \ref{lemma1}}. \ \ (1) It is a a consequence of Theorem \ref{th21}. (2) Note that
\begin{align*}
&\e|\rho_m'(e_t)-\psi(e_t)|=\int |\rho_m'(u)-\psi(u)|f(u)du\\
=&\int \left|\int \rho'(x+u)\phi_m(x)dx-\rho'(u)\right|f(u)du\\
\le&\iint |\rho'(x+u)-\rho'(u)|\phi_m(x)dxf(u)du\\
=&\int \left[\int |\rho'(x+u)-\rho'(u)|f(u)du\right]\phi_m(x)dx\\
=&\int_0^\infty \left[\int (\rho'(x+u)-\rho'(u))f(u)du\right]\phi_m(x)dx\\
&+\int_{-\infty}^0 \left[\int (\rho'(u)-\rho'(x+u))f(u)du\right]\phi_m(x)dx\\
=&-\int_0^\infty \left[\int (\rho(x+u)-\rho(u))f'(u)du\right]\phi_m(x)dx\\
&-\int_{-\infty}^0 \left[\int (\rho(u)-\rho(x+u))f'(u)du\right]\phi_m(x)dx\\
\le&\int \left[\int |\rho(u)-\rho(x+u)||f'(u)|du\right]\phi_m(x)dx,
\end{align*}
by integration by parts and noting that $\rho'(u)=\psi(u)$ is an increasing function. It follows that
\begin{align*}
\e|\rho_m'(e_t)-\psi(e_t)|
\le C\int |x|\phi_m(x)dx \int |f'(u)|du=Cm^{-1}.
\end{align*}

(3) Note that
\begin{align*}
&\e[\rho_m''(e_t)-\rho''(e_t)]=\int [\rho_m''(u)-\rho''(u)]f(u)du\\
=&\iint [\rho''(x+u)-\rho''(u)]\phi_m(x)dxf(u)du\\
=&\int \left[\int [\rho''(x+u)-\rho''(u)]f(u)du\right]\phi_m(x)dx\\
=&\int \left[\int [\rho(x+u)-\rho(u)]f''(u)du\right]\phi_m(x)dx,
\end{align*}
which implies $|\e[\rho_m''(e_t)-\rho''(e_t)]|\le C m^{-1}$ immediately.

(4) Note that
\begin{align*}
& |\e[\rho''(e_t+\epsilon)-\rho''(e_t)]|
=\left|\int[\rho''(u+\epsilon)-\rho''(u)]f(u)du\right|\\
=&\left|\int[\rho(u+\epsilon)-\rho(u)]f''(u)du\right|
\le\int|\rho(u+\epsilon)-\rho(u)||f''(u)|du\\
\le& |\epsilon| \int |f''(u)|du=C|\epsilon|.
\end{align*}
Therefore,
\begin{align*}
|\e[\rho_m''(e_t+\epsilon)-\rho''_m(e_t)]|\le& |\e[\rho''(e_t+\epsilon)-\rho''(e_t)]|+|\e[\rho_m''(e_t+\epsilon)-
\rho''(e_t+\epsilon)]|\\
&+|\e[\rho_m''(e_t)-\rho''(e_t)]|\\
=&O(m^{-1}+\epsilon).
\end{align*}

{\bf Proof of Theorem \ref{th22}}.\ \ Define
\begin{equation*}
\tilde{L}_{mn}(\beta)\equiv \sum_{i=1}^n[\rho_m(e_i- n^{-1/2} x_i^\top\beta)-\rho_m(e_i)].
\end{equation*}

It is readily seen from Theorem \ref{th21}, $\sup_u|\rho_m(u)-\rho(u)|\le C m^{-1}$, that
\begin{equation}\label{2c}
\sup_{\beta\in \mathbb{R}^d} \left|\tilde{L}_n(\beta)-\tilde{L}_{mn}(\beta)\right|
=O(nm^{-1}), \ \ w.p.1.
\end{equation}
%In the sequel let $m=[n^{1+\varepsilon}]$ with $\varepsilon>0$.

Now, using Taylor expansion we have
\begin{align*}
\rho_m(e_i-n^{-1/2}x_i^\top\beta)-\rho_m(e_i)
 =&-\DF{1}{\sqrt{n}}\rho'_m(e_i)x_i^\top\beta+
 \DF{1}{2n}\rho_m''(e_i)[x_i^\top\beta]^2\\
 &- \DF{1}{6n\sqrt{n}}(x_i^\top\beta)^3
 \int_0^1\rho_m'''(e_i-wn^{-1/2}x_i^\top\beta)dw.
\end{align*}
Recall that
\begin{equation*}
\rho_m'''(u)=-\int \rho(x)\phi_m'''(x-u)dx=-\int \rho(x)m^4\phi'''(m(x-u))dx
\end{equation*}
and due to Assumptions 2.1(a) and 2.2(c), $\rho(x)m^4\phi'''(m(x-u))\to 0$ uniformly in $x\ne u$ as $m\to\infty$, whereas for $x=u$, $\phi'''(m(x-u))=0$ since $\phi(\cdot)$ is symmetric. It follows that the remainder term in the Taylor expansion is $O_P(n^{-3/2})$ and therefore is negligible in the sequel.

Accordingly, we may write
\begin{align*}
\tilde{L}_{mn}(\beta)=&\sum_{t=1}^n\left(-\DF{1}{\sqrt{n}}
\rho'_m(e_t)x_t^\top\beta
+\DF{1}{2n}\rho_m''(e_t)[x_t^\top\beta]^2\right) \\
=&\left(-\DF{1}{\sqrt{n}}\sum_{t=1}^n\rho'_m(e_t)x_t^\top\right)\beta
+\DF{1}{2}\beta^\top \left(\DF{1}{n} \sum_{t=1}^n\rho''_m(e_t)x_tx_t^\top\right)\beta \\
=&\left(-\DF{1}{\sqrt{n}}\sum_{t=1}^n\psi(e_t)x_t^\top\right)\beta+\DF{1}{2} \beta^\top \left(\DF{1}{n} \sum_{t=1}^n\e[\rho''_m(e_t)]x_tx_t^\top\right)\beta \\
&+\left(\DF{1}{\sqrt{n}}\sum_{t=1}^n[\rho'_m(e_t)-\psi(e_t)]x_t^\top\right)\beta \\
&+\DF{1}{2}\beta^\top \left(\DF{1}{n} \sum_{t=1}^n\{\rho''_m(e_t)-\e[\rho''_m(e_t)]\} x_tx_t^\top\right)\beta  \\
=&\left(-\DF{1}{\sqrt{n}}\sum_{t=1}^n\psi(e_t)x_t^\top\right)\beta+\DF{a}{2}
\beta^\top \left(\DF{1}{n} \sum_{t=1}^nx_tx_t^\top\right)\beta \\
&+\left(\DF{1}{\sqrt{n}}\sum_{t=1}^n[\rho'_m(e_t)-\psi(e_t)]x_t^\top\right)\beta \\
&+\DF{1}{2}\beta^\top \left(\DF{1}{n} \sum_{t=1}^n\{\rho''_m(e_t)-\e[\rho''_m(e_t)]\} x_tx_t^\top\right)\beta \\
&+\DF{1}{2}(\e[\rho''_m(e_t)]-a)
\beta^\top \left(\DF{1}{n} \sum_{t=1}^nx_tx_t^\top\right)\beta
\end{align*}
where $a=\e[\rho''(e_t)]>0$ due to convexity. Moreover, due to Lemma \ref{lemma1}, $\e|\rho'_m(e_t)-\psi(e_t)|\le Cm^{-1}$, and $\e[\rho''_m(e_t)-a]=O(m^{-1})$. Accordingly,
\begin{align}
&\DF{1}{\sqrt{n}}\sum_{t=1}^n[\rho'_m(e_t)-\psi(e_t)]x_t=O_P(n^{1/2}m^{-1})
\label{pth22a}\\
 &\DF{1}{n} \sum_{t=1}^n\{\rho''_m(e_t) -\e[\rho''_m(e_t)]\} x_tx_t^\top=O_P(n^{-1/2}).\label{pth22b}
\end{align}
The first assertion \eqref{pth22a} is readily seen due to Lemma \ref{lemma1}, while we give the proof of \eqref{pth22b} below. Indeed,
\begin{align*}
&\e\left\|\DF{1}{n} \sum_{t=1}^n\{\rho''_m(e_t)-\e[\rho''_m(e_t)]\} x_tx_t^\top\right\|^2\\
=&\DF{1}{n^2} \sum_{t=1}^n\e\{\rho''_m(e_t)-\e[\rho''_m(e_t)]\}^2\e \|x_t\|^4\\
=&C\DF{1}{n}\{\e[\rho''_m(e_t)]^2-(\e[\rho''_m(e_t)])^2\},
\end{align*}
and
\begin{align*}
\e[\rho''_m(e_t)]^2=&\int [\rho''_m(u)]^2f(u)du
= \int \left[\int \rho(x)\phi''_m(x-u)dx\right]^2f(u)du\\
=&\int \left[\int \rho(x+u)\phi''_m(x)dx\right]^2f(u)du\\
=&\int \left[\int \rho(x+u)m\sqrt{\DF{m}{\pi}}(2mx^2-1)e^{-mx^2}dx\right]^2f(u)du\\
=&\int \left[\int \rho\left(\DF{x}{\sqrt{m}}+u\right)m \DF{1}{\sqrt{\pi}}(2x^2-1)e^{-x^2}dx\right]^2f(u)du\\
=&\int \left[\int \left(\rho''(u)\DF{x^2}{m}+O(m^{-2})\right)m \DF{1}{\sqrt{\pi}}(2x^2-1)e^{-x^2}dx\right]^2f(u)du\\
=&\int \left[\rho''(u)\int \DF{1}{\sqrt{\pi}} x^2(2x^2-1)e^{-x^2}dx\right]^2f(u)du+O(m^{-1})\\
=&\int[\rho''(u)]^2f(u)du+O(m^{-1}),
\end{align*}
by Taylor expansion and integrals $\int x^j(2x^2-1)e^{-x^2}dx=0$, $j=0,1,3$, and similarly
\begin{align*}
\e\rho''_m(e_t)=&\int \rho''_m(u)f(u)du=\iint \rho(x+u)\phi''_m(x)dxf(u)du \\
 =& \iint \rho(x+u)m\sqrt{\DF{m}{\pi}}(2mx^2-1)e^{-mx^2}dx f(u)du\\
 =&\iint \rho\left(\DF{x}{\sqrt{m}}+u\right)m \DF{1}{\sqrt{\pi}}(2x^2-1)e^{-x^2}dxf(u)du\\
 =&\iint \left(\rho''(u)\DF{x^2}{m}+O(m^{-2})\right)m \DF{1}{\sqrt{\pi}}(2x^2-1)e^{-x^2}dxf(u)du\\
 =&\int \rho''(u)f(u)du+O(m^{-1}).
\end{align*}
Thus,
\begin{equation*}
\e[\rho''_m(e_t)]^2-(\e[\rho''_m(e_t)])^2=\int[\rho''(u)]^2f(u)du-\left(\int \rho''(u)f(u)du\right)^2+O(m^{-1}).
\end{equation*}

It is readily seen that $|\tilde{L}_{mn}(\beta)-Q_n(\beta)|=O_P(n^{-1/2}+n^{1/2}m^{-1})$, that together with \eqref{2c} and $m=n^{\epsilon+3/2}$ (any $\epsilon>0$) implies \eqref{th22b}.

Moreover,
\begin{align}\label{2f}
&\sup_{\|\beta\|\le c}|\tilde{L}_{mn}(\beta)-Q_n(\beta)|\notag\\
\le&\sup_{\|\beta\|\le c}
\left|\left(\DF{1}{\sqrt{n}}\sum_{t=1}^n[\rho'_m(e_t)-\psi(e_t)]x_t^\top\right)\beta \right|\notag\\
&+\sup_{\|\beta\|\le c}
\left|\DF{1}{2}\beta^\top \left(\DF{1}{n} \sum_{t=1}^n\{\rho''_m(e_t)-\e[\rho''_m(e_t)]\} x_tx_t^\top\right)\beta \right|\notag \\
&+\sup_{\|\beta\|\le c}
\left|\DF{1}{2}(\e[\rho''_m(e_t)]-a_2)
\beta^\top \left(\DF{1}{n} \sum_{t=1}^nx_tx_t^\top\right)\beta\right|\notag\\
\le&c\left\|\DF{1}{\sqrt{n}}\sum_{t=1}^n[\rho'_m(e_t)-\psi(e_t)]x_t \right\|
+\DF{c^2}{2}\left\|\DF{1}{n} \sum_{t=1}^n\{\rho''_m(e_t)-\e[\rho''_m(e_t)]\} x_tx_t^\top \right\|\notag \\
&+\DF{c^2}{2}|\e[\rho''_m(e_t)]-a_2|\left\|\DF{1}{n} \sum_{t=1}^n x_tx_t^\top \right\|\notag \\
=&O_P(m^{-1}n^{1/2}+n^{-1/2}+m^{-1})=O_P(n^{-1/2}),
\end{align}
for any $c>0$.

It is also clear that
\begin{equation}\label{2g}
\sup_{\|\beta\|\le c_n}|\tilde{L}_{mn}(\beta)-Q_n(\beta)|=o_P(1),
\end{equation}
for $c_n=o(n^{1/4})$.

In view of \eqref{2c}, along with \eqref{2f} and \eqref{2g}, assertions \eqref{th22c} and \eqref{th22d} follow immediately.

For simplicity in the sequel, denote
\begin{equation*}
S_n=\DF{1}{\sqrt{n}}\sum_{t=1}^n\psi(e_t)x_t, \ \ \Sigma_n=
\DF{a}{n} \sum_{t=1}^nx_tx_t^\top,
\end{equation*}
so that $Q_n(\beta)=-S_n^\top\beta+\beta^\top\Sigma_n\beta/2$ and $Q_n(\beta)$ has unique minimizer $\widehat{\beta}_Q=\Sigma_n^{-1}S_n$ that converges in distribution given that $\e[\psi(e_t)]=0$.

Thus, we may write
\begin{align}\label{qnre}
Q_n(\beta)=&-S_n^\top\beta+\DF{1}{2}\beta^\top\Sigma_n\beta\notag \\
=&\DF{1}{2}(\beta-\widehat{\beta}_Q)^\top\Sigma_n(\beta-\widehat{\beta}_Q)\notag\\
&-S_n^\top\beta+\DF{1}{2}\beta^\top\Sigma_n\beta-\DF{1}{2}
(\beta-\widehat{\beta}_Q)^\top
\Sigma_n(\beta-\widehat{\beta}_Q)\notag\\
=&\DF{1}{2}(\beta-\widehat{\beta}_Q)^\top\Sigma_n(\beta-\widehat{\beta}_Q)
-\DF{1}{2}\widehat{\beta}_Q^\top\Sigma_n\widehat{\beta}_Q.
\end{align}
This reformulation will be used later.

We next shall show
\begin{equation}\label{2h}
\widehat{\beta}-\widehat{\beta}_Q=o_P(d_n), \ \ \ n\to\infty,
\end{equation}
where $\widehat{\beta}$ is given by \eqref{2b} and $d_n=n^{-1/4}\log\log(n)$. To the end, for each $\epsilon>0$, we shall show $\p(\|\widehat{\beta} -\widehat{\beta}_Q\|>d_n\epsilon)\to 0$ as $n\to\infty$.

Denote $B_n=\{\beta:\ \|\beta-\widehat{\beta}_Q\|\le d_n\epsilon\}$ a ball with center $\widehat{\beta}_Q$ and radius $d_n\epsilon$. Because $\widehat{\beta}_Q$ is bounded in probability, $B_n$ will be covered by some compact set, $\|\beta\|\le c$ say, and then \eqref{th22c} implies that
\begin{equation*}
\sup_{\beta\in B_n}|\tilde{L}_{n}(\beta)-Q_n(\beta)|\le \sup_{\|\beta\|\le c } |\tilde{L}_{n}(\beta)-Q_n(\beta)|=O_P(n^{-1/2}).
\end{equation*}

Denote $R_n(\beta)\equiv \tilde{L}_{n}(\beta)-Q_n(\beta)$ and $r_n\equiv\sup_{\beta\in B_n}|R_n(\beta)|=O_P(n^{-1/2})$. If $\beta\not\in B_n$, $\beta=\widehat{\beta}_Q+\lambda d_nv$ where $v$ is a unit vector and $\lambda>\epsilon$. Let $\beta^*=\widehat{\beta}_Q+\epsilon d_nv$ be the boundary point on the segment joining $\widehat{\beta}_Q$ and $\beta$.

It follows by the convexity of $\tilde{L}_n(\cdot)$ and \eqref{qnre} that
\begin{align*}
&\DF{\epsilon}{\lambda}\tilde{L}_n(\beta)+\left(1-\DF{\epsilon}{\lambda}\right)
\tilde{L}_n(\widehat{\beta}_Q)\ge \tilde{L}_n(\beta^*)\\
=&Q_n(\beta^*)+R_n(\beta^*)\\
\ge&\DF{1}{2}\epsilon^2d_n^2 v^\top\Sigma_nv- \DF{1}{2}\widehat{\beta}_Q^\top\Sigma_n\widehat{\beta}_Q-r_n\\
\ge&\DF{1}{2}\epsilon^2d_n^2a\lambda_{\min}+
\tilde{L}_n(\widehat{\beta}_Q)-2r_n,
\end{align*}
where $\lambda_{\min}>0$ is the minimum eigenvalue of $\Sigma_n$ apart from the factor $a$. Rearranging to have
\begin{align*}
\inf_{\|\beta-\widehat{\beta}_Q\|>d_n\epsilon}\tilde{L}_n(\beta)\ge & \tilde{L}_n(\widehat{\beta}_Q)+d_n^2\DF{\lambda}{2\epsilon}\left[
\epsilon^2a\lambda_{\min}-4d_n^{-2}r_n\right]\\
\ge &\tilde{L}_n(\widehat{\beta}_Q)+d_n^2\DF{\lambda}{4}
\epsilon a\lambda_{\min},
\end{align*}
since $d_n^{-2}r_n=o_P(1)$ and eventually $4d_n^{-2}r_n<\epsilon^2a \lambda_{\min}/2$ in probability. As a result, it is impossible that $\tilde{L}_n(\beta)$ attains its minimum beyond $\|\beta-\widehat{\beta}_Q\| >d_n\epsilon$, that is, $\|\widehat\beta-\widehat{\beta}_Q\|\le d_n\epsilon$ in probability. This proves \eqref{2h}. \qed

}

\end{document}